\documentclass[aps,prb,superscriptaddress,twocolumn,floatfix]{revtex4-1}

\usepackage{graphicx}
\usepackage{hyperref}
\usepackage{threeparttable}

\begin{document}

\title{Electronic structure of \textit{R}Sb (\textit{R} = Y, Ce, Gd, Dy, Ho, Tm, Lu) studied by angle-resolved photoemission spectroscopy}

\author{Yun Wu}
\affiliation{Division of Materials Science and Engineering, Ames Laboratory, Ames, Iowa 50011, USA}
\affiliation{Department of Physics and Astronomy, Iowa State University, Ames, Iowa 50011, USA}

\author{Yongbin Lee}
\affiliation{Division of Materials Science and Engineering, Ames Laboratory, Ames, Iowa 50011, USA}

\author{Tai Kong}
\altaffiliation[Present Address: ]{Department of Chemistry, Princeton University, Princeton, NJ, 08544, USA}

\author{Daixiang Mou}
\author{Rui Jiang}
\author{Lunan Huang}
\author{S.~L.~Bud'ko}
\author{P. C. Canfield}
\email[]{canfield@ameslab.gov}
\author{Adam Kaminski}
\email[]{kaminski@ameslab.gov}
\affiliation{Division of Materials Science and Engineering, Ames Laboratory, Ames, Iowa 50011, USA}
\affiliation{Department of Physics and Astronomy, Iowa State University, Ames, Iowa 50011, USA}

\date{\today}

\begin{abstract}
We use high resolution angle-resolved photoemission spectroscopy (ARPES) and electronic structure calculations to study the electronic properties of rare-earth monoantimonides \textit{R}Sb (\textit{R} = Y, Ce, Gd, Dy, Ho, Tm, Lu). The experimentally measured Fermi surface (FS) of \textit{R}Sb consists of at least two concentric hole pockets at the $\Gamma$ point and two intersecting electron pockets at the $X$ point. These data agree relatively well with the electronic structure calculations. Detailed photon energy dependence measurements using both synchrotron and laser ARPES systems indicate that there is at least one Fermi surface sheet with strong three-dimensionality centered at the $\Gamma$ point. Due to the ``lanthanide contraction'', the unit cell of different rare-earth monoantimonides shrinks when changing rare-earth ion from CeSb to LuSb. This results in the differences in the chemical potentials in these compounds, which is demonstrated by both ARPES measurements and electronic structure calculations. Interestingly, in CeSb, the intersecting electron pockets at the $X$ point seem to be touching the valence bands, forming a four-fold degenerate Dirac-like feature. On the other hand, the remaining rare-earth monoantimonides show significant gaps between the upper and lower bands at the $X$ point. Furthermore, similar to the previously reported results of LaBi, a Dirac-like structure was observed at the $\Gamma$ point in YSb, CeSb, and GdSb, compounds showing relatively high magnetoresistance. This Dirac-like structure may contribute to the unusually large magnetoresistance in these compounds. 
\end{abstract}

\maketitle

\section{Introduction}

Rare-earth monoantimonides \textit{R}Sb (\textit{R}=rare-earth) have attracted great attention due to their remarkable magnetic and electronic properties~\cite{Child63PhysRev, Busch65PhysLett, Cooper70PRB, Rossat77PRB, Duan2007Electronic}. Although these compounds crystallize in the simple NaCl-type cubic structure~\cite{Child63PhysRev}, most of them exhibit strongly anisotropic magnetic properties below their N$\acute{\text{e}}$el Temperatures~\cite{Busch65PhysLett, Busch68JApplPhys, Cooper70PRB}. TbSb, HoSb, and ErSb become antiferromagnetic at low temperature, showing the MnO-type arrangement of magnetic moments, i.e., with ferromagnetic sheets perpendicular to the cube diagonal, and magnetic moments in adjacent sheets anti-parallelly arranged~\cite{Child63PhysRev}. Most of the \textit{R}Sb (except for GdSb) studied by Busch \textit{et al}~\cite{Busch65PhysLett} show metamagnetic properties, i.e., the spin structure changes abruptly from antiferromagnetism to a spin arrangement with a net magnetic moment under sufficiently high magnetic fields. Further studies~\cite{Busch68JApplPhys} of these compounds show that strong anisotropy is found in the monoantimonides of Ce, Nd, Dy, and Ho (in agreement with the Ising model), whereas TbSb and ErSb only exhibit weak anisotropy. DySb has been shown to have a single first-order magnetic phase transition through specific-heat, susceptibility, and neutron scattering measurements~\cite{Bucher72PRL}. Among these compounds, CeSb has the most complicated magnetic phase diagram with at least 14 distinct metamagnetic states at low temperatures and magnetic fields~\cite{Rossat77PRB, Wiener00JAC}. In CeSb, the largest observable Kerr rotation (90 ${}^{\circ}$) in a single reflection has also been reported~\cite{Pittini1996Discovery}. Recently, extremely large magnetoresistance~\cite{Mun12PRB, Ali14Nat, Liang15NatMat, Narayanan2015Linear} has attracted tremendous attention. Not only do the materials with this type of property have potential applications such as magnetic field sensors, but also are platforms for studying exotic physical properties, such as Dirac node arc states~\cite{Wu2016Dirac}, type-II Weyl fermion states~\cite{Soluyanov2015Type, Bruno2016Surface, Wang2016Observation, Wu2016Observation, Feng2016Spin}, three-dimensional Dirac states~\cite{Wang13PRB, Neupane14NatCom, Liu14NatMat, Borisenko14PRL}, etc. Interestingly, CeSb also shows relatively high magnetoresistance of 9000~\% at 5~K and 5.5~T~\cite{Wiener00JAC}. Besides CeSb, GdSb show even higher magnetoresistance, reaching $1.25 \times {10}^{4}$ \% at 4.2~K and 10~T~\cite{Li96PRB}. All the researches indicate that different rare-earth elements would have different impacts on the electronic and magnetic properties of these compounds. If we are measuring the electronic properties of these compounds at $T > {T}_{N}$, the different ionic sizes (due to lanthanide contraction) may have a significant effect on the electronic structure of these materials. Thus, in order to understand the role that lanthanide contraction plays in these compounds, detailed electronic structure measurements of \textit{R}Sb are necessary.

A number of electronic properties of \textit{R}Sb were previously studied using band structure calculations~\cite{Hasegawa77JPSJ, Hasegawa80JPC, Hasegawa85JPSJ, Liechtenstein94PRB}, quantum oscillations~\cite{Kitazawa88JMMM, Settai94JPSJ}, and ARPES measurements~\cite{Olson96PRL, Kumigashira97PRB, Kumigashira98PRB, Kumigashira99PhysB, Kumigashira00PRB, Takayama09JPSJ}. However, systematic ARPES studies of the rare-earth monoantimonides, especially photon energy dependent measurements, are still needed to better understand these materials. Here, we present the study of Fermi surface and band dispersion of \textit{R}Sb (\textit{R}=Y, Ce, Gd, Dy, Ho, Tm, Lu), with specific emphasis on their 3D character, using high resolution synchrotron and tunable VUV laser ARPES measurements. The FS of \textit{R}Sb consists of at least two hole pockets at the $\Gamma$ point and two, intersecting, electron pockets at the $X$ point. We also determined the band structure at the $\Gamma$ point along the out of plane (${k}_{z}$) direction, which shows strong three-dimensionality. Interestingly, a four-fold degenerate Dirac-like feature was observed at the $X$ point in CeSb, consistent with the previously reported results~\cite{Alidoust2016New}. However, other compounds, such as GdSb and YSb, show significant gaps between the conduction and valence bands at the $X$ point. Furthermore, a Dirac-like feature is observed at the $\Gamma$ point within specific photon energy range in YSb, CeSb, and GdSb, which may contribute to the unusually high magnetoresistance observed in these compounds.

\section{Experimental details}

\begin{table}
\begin{threeparttable}
\caption{Physical properties of \textit{R}Sb (\textit{R} = Rare earth)}
\begin{tabular}{c | c | c | c}
	\hline
	\hspace{3mm}\textit{R}Sb \hspace{3mm}	& \hspace{1mm} lattice a (\AA)~\tnote{I}	\hspace{1mm} & \hspace{1mm} ionic radii (\AA)~\tnote{II} \hspace{1mm}	& \hspace{2mm} ${T}_{N}$ (K)~\tnote{III} \hspace{2mm} \\
	\hline
	YSb				& 6.190		& 0.9~\tnote{IV}		& 				\\
	CeSb			& 6.408		& 1.01				& 16.7~\tnote{V}	\\
	GdSb			& 6.210		& 0.938 			& 28	 			\\
	DySb			& 6.150		& 0.912				& 9.5~\tnote{VI} 	\\
	HoSb			& 6.130		& 0.901				& 5.5~\tnote{VII}	\\
	TmSb			& 6.090		& 0.88				& 					\\
	LuSb			& 6.060		& 0.86				& 		 			\\
	\hline
\end{tabular}
\begin{tablenotes}\footnotesize
\item[I, II, III, IV, V] Data from Ref.~\onlinecite{Abdusalyamova1990Investigation, Jia1991Crystal, Busch65PhysLett, ShannonRadii, Wiener00JAC}. \item[VI] Ref.~\onlinecite{Nereson1972Neutron} reported a value of 12~K. \item[VII] Ref.~\onlinecite{Child63PhysRev} reported a value of 9~K.
\end{tablenotes}

\label{tab:Tab1}
\end{threeparttable}
\end{table}

\begin{figure*}[tbp]
	\includegraphics[width=6.5in]{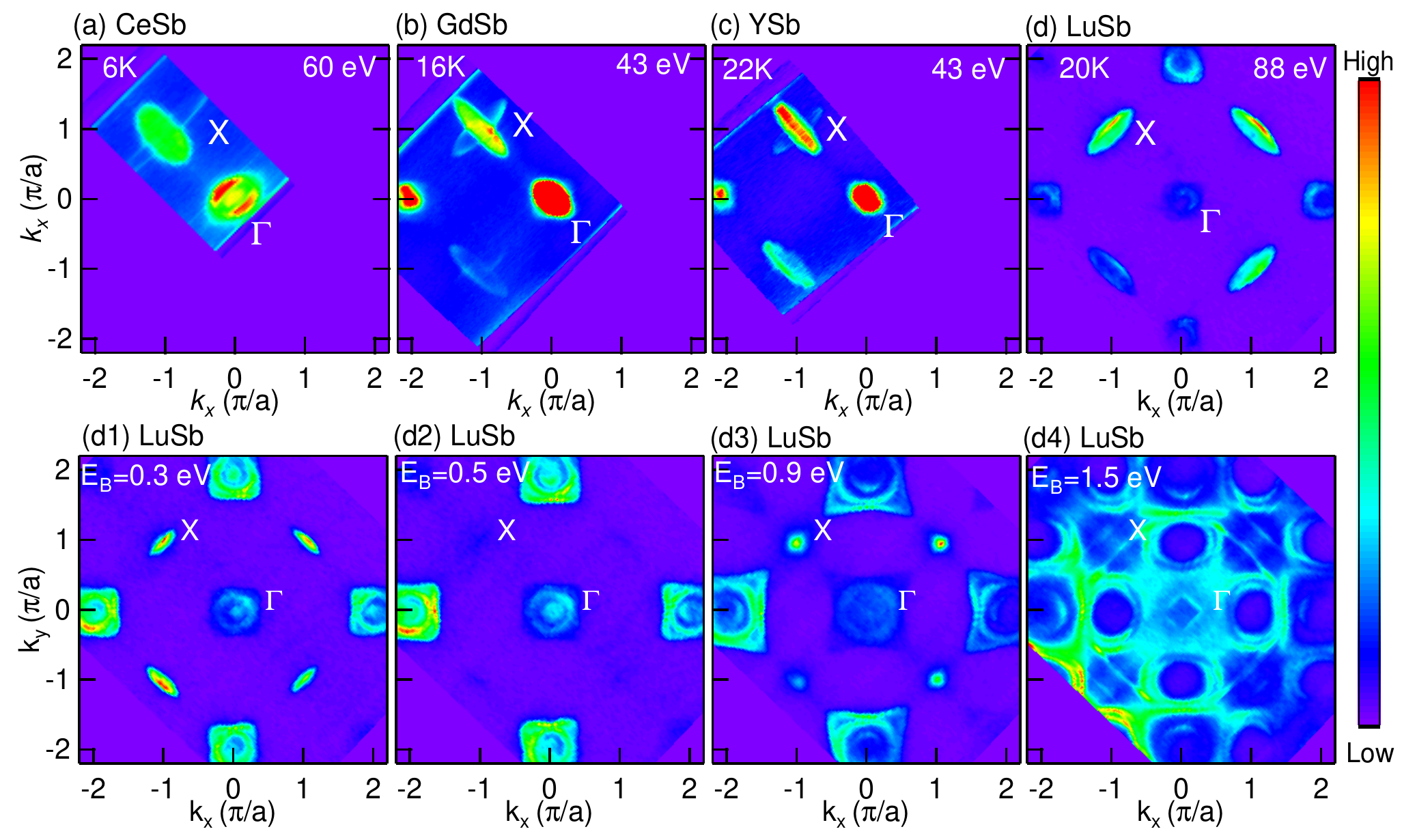}%
	\caption{Constant energy contour plots of \textit{R}Sb (\textit{R} = Ce, Gd, Y, and Lu).
	(a)--(d) Fermi surface plots of ARPES intensity integrated within 10 meV about the chemical potential, corresponding to CeSb, GdSb, YSb, and LuSb, respectively. The specific temperature and incident photon energy used during the measurements are marked at the top left and right corners, respectively. Red dashed lines in (b) mark the outline of the two intersecting electron pockets.
	(d1)--(d4) Constant energy contour plots of LuSb measured using the photon energy of 88~eV at the binding energies of 0.3, 05, 0.9, and 1.5~eV, respectively. }
	\label{fig:Fig1}
\end{figure*}

\begin{figure*}[tbp]
	\includegraphics[width=6.5in]{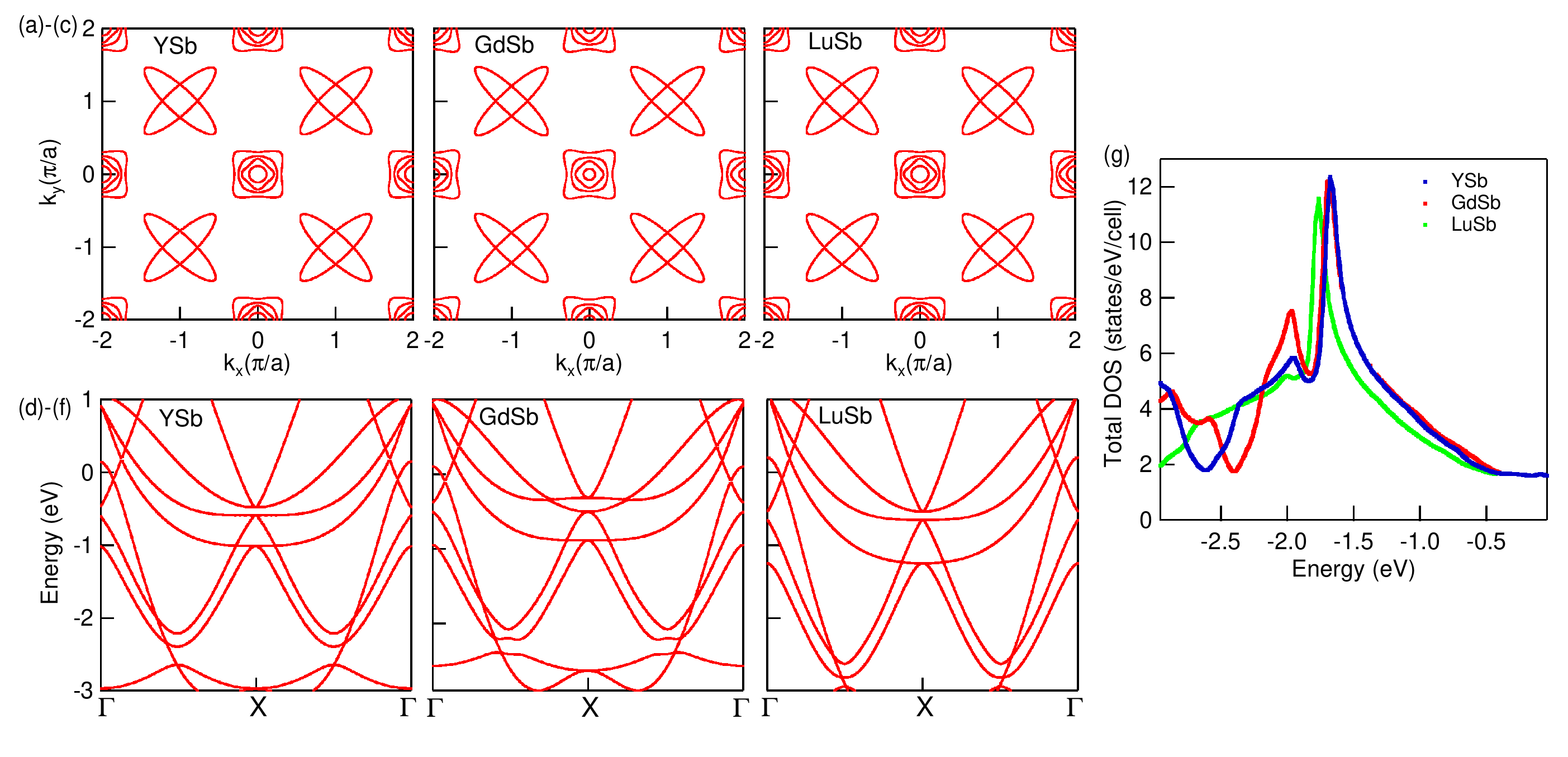}%
	\caption{Calculated Fermi surface and band dispersion of YSb, GdSb, and LuSb.
	(a)--(c) Calculated Fermi Surface of YSb, GdSb, and LuSb.
	(d)--(f) Calculated Band structure along $\Gamma-X$ of YSb, GdSb, and LuSb.
	(g) Calculated Density of states (DOS) of YSb, GdSb, and LuSb}
	\label{fig:Fig2}
\end{figure*}

\begin{figure*}[bt]
	\includegraphics[width=6.5in]{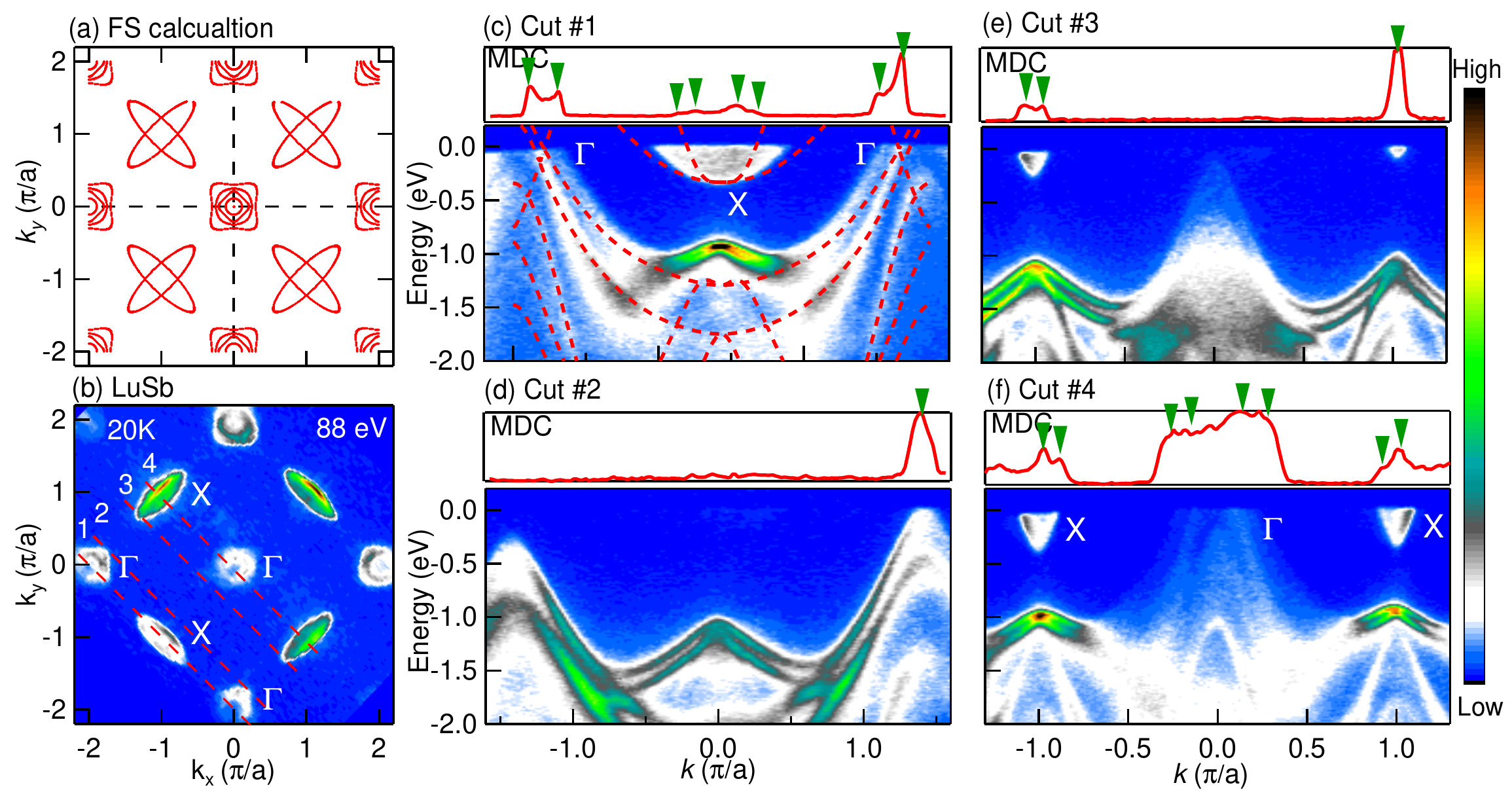}%
	\caption{Fermi surface and band dispersion of LuSb measured at $T=$20~K and photon energy of 88~eV.
	(a) Calculated Fermi Surface using FPLAPW method.
	(b) Fermi surface plot of ARPES intensity integrated within 10 meV about the chemical potential of LuSb measured at $T=$20~K and photon energy of 88~eV.
	(c)--(f) Band dispersion along cuts 1--4 in panel (b). The top panels show the Momentum Dispersion Curves (MDCs) at the Fermi level. The green arrows point to the obvious Fermi crossings. The red dashed lines in panel (c) are the results from band structure calculations.
	\label{fig:Fig3}}
\end{figure*}

\begin{figure*}[tb]
	\includegraphics[width=6.5in]{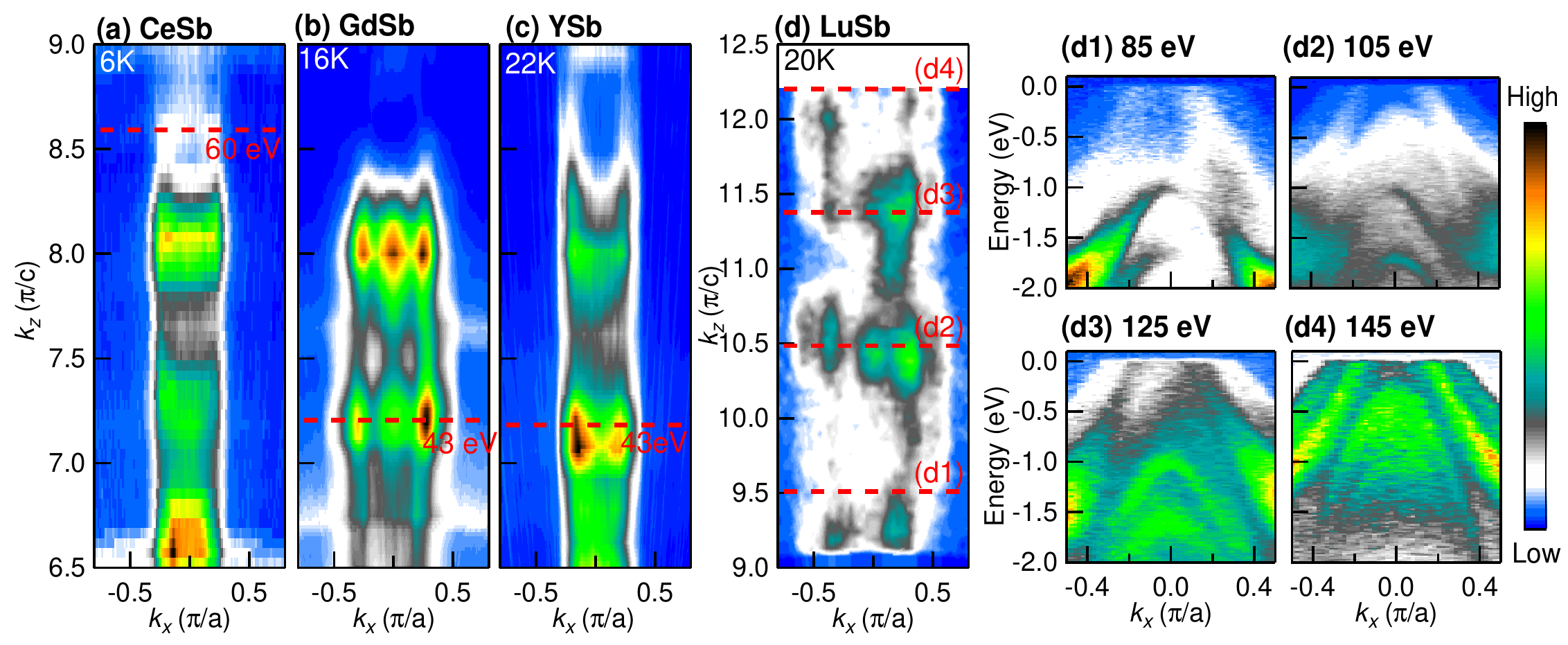}
	\caption{Out of plane momentum (${k}_{z}$) dispersion of YSb, CeSb, GdSb, and LuSb at the $\Gamma$ point measured using photon energies in the 20--150~eV range. The corresponding temperatures during the measurements are marked at the top left corner in each plot.
	(a) ${k}_{z}$ dispersion of CeSb measured at SRC using photon energies in the 30 to 80~eV range with 2~eV step.
	(b) ${k}_{z}$ dispersion of GdSb measured at ALS using photon energies in the 20 to 80~eV range with 1~eV step.
	(c) ${k}_{z}$ dispersion of YSb measured at SRC using photon energies in the 30 to 76~eV range with 2~eV step.
	(d) ${k}_{z}$ dispersion of LuSb measured at ALS using photon energies in the 77 to 145~eV range with 1~eV step.
	(d1)--(d4) band dispersion of LuSb at the $\Gamma$ point measured using the corresponding incident photon energies.
	\label{fig:Fig4}}
\end{figure*}

\begin{figure*}[tb]
	\includegraphics[width=5in]{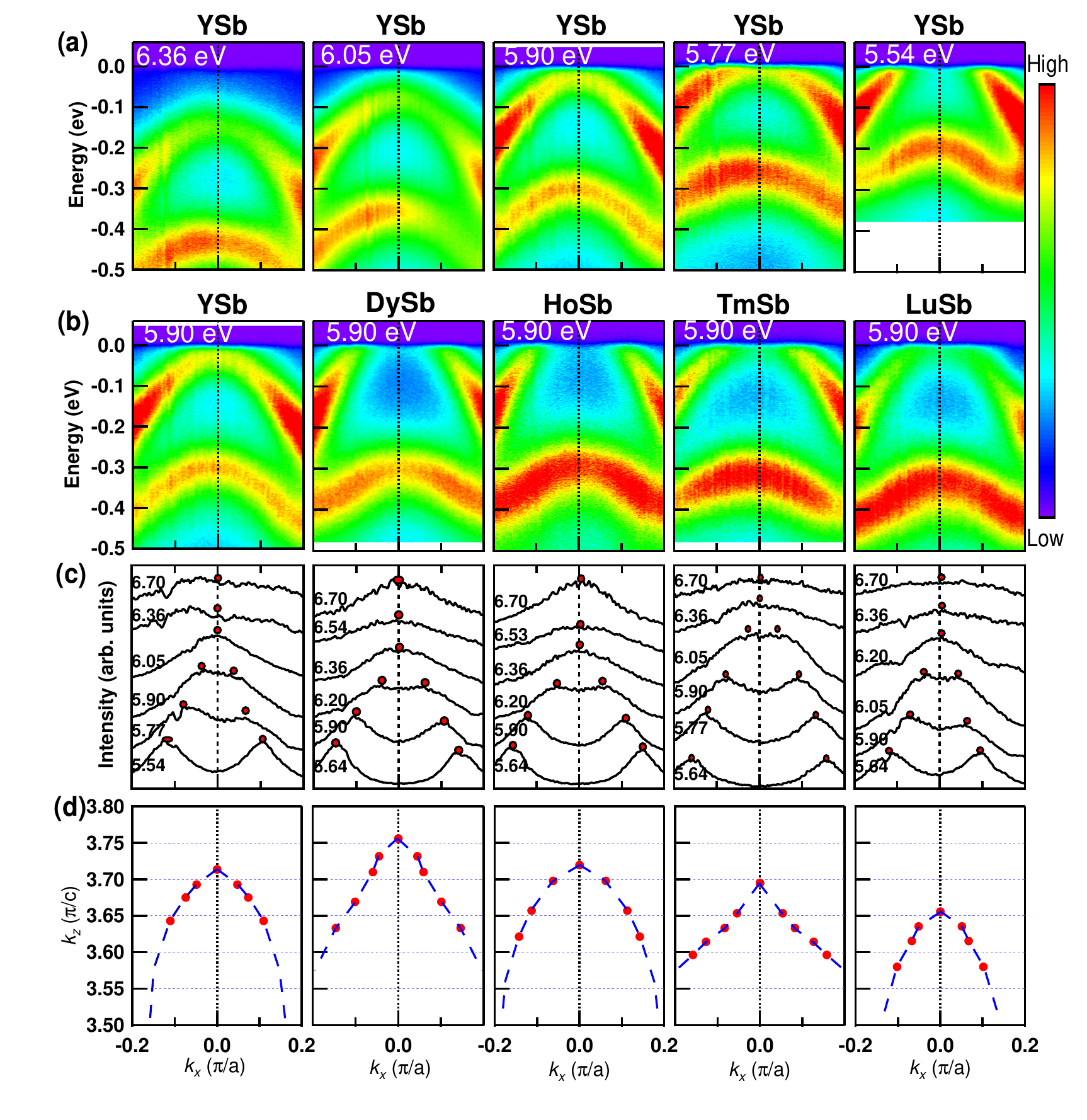}
	\caption{Band dispersion of \textit{R}Sb at the $\Gamma$ point measured at $T=$40~K and using different photon energies. 
	(a) band dispersion of YSb measured using various photon energies marked at the top of each plot. 
	(b) band dispersion of \textit{R}Sb (\textit{R} = Y, Dy, Ho, Tm and Lu, from left to right, respectively) measured
	using the photon energy of 5.90~eV. 
	(c) MDCs at ${E}_{F}$ measured using different photon energies. The photon energies are marked on the left hand side of the corresponding MDCs. The red solid dots mark the peak positions of the MDCs. 
	(d) ${k}_{z}$ dispersion of \textit{R}Sb (\textit{R} = Y, Dy, Ho, Tm and Lu, from left to right, respectively) extracted from panel (c). The red solid dots are reproduced from panel (c) with each photon energy marked on the left hand side of the MDCs. The blue dashed lines are guides to the eye.
	\label{fig:Fig5}}
\end{figure*}

\begin{figure*}[tb]
	\includegraphics[width=5in]{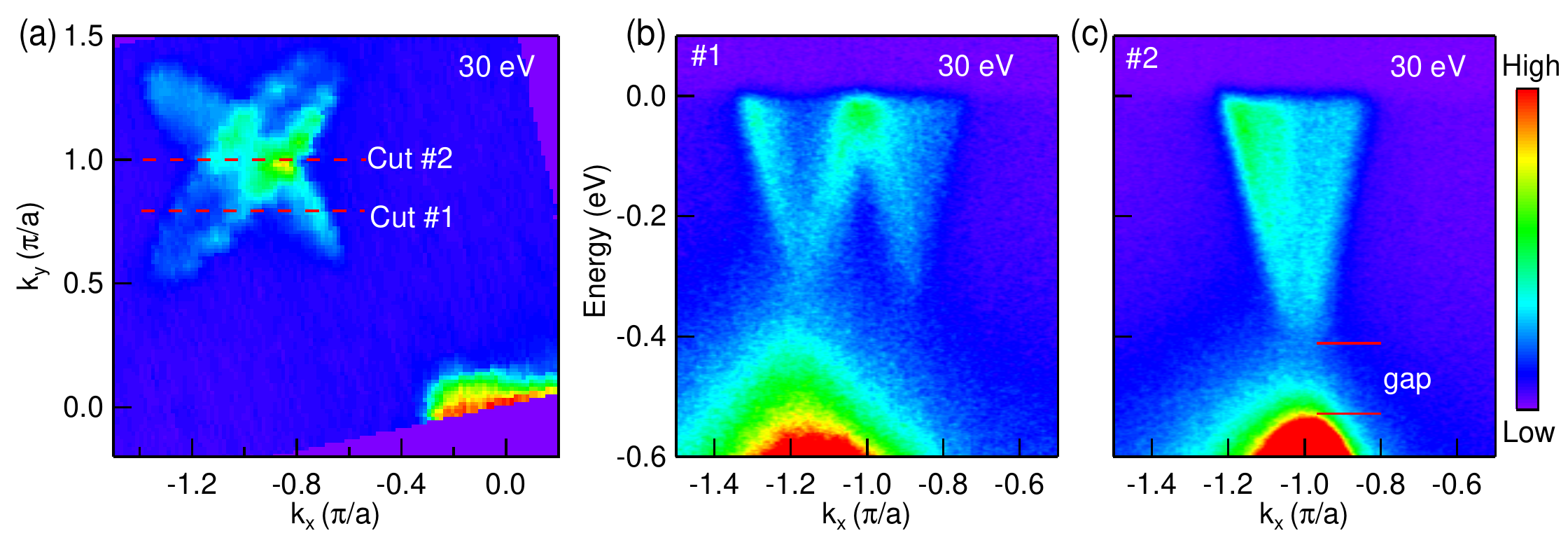}%
	\caption{Fermi surface plot and band dispersion of GdSb measured at $T=35$~K using the photon energy of 30~eV.
	(a) Fermi surface plot of ARPES intensity integrated within 10 meV about the chemical potential. 
	(b)--(c) Band dispersion along cuts 1--2.
	\label{fig:Fig6}}
\end{figure*}

\begin{figure*}[bt]
	\includegraphics[width=6.5in]{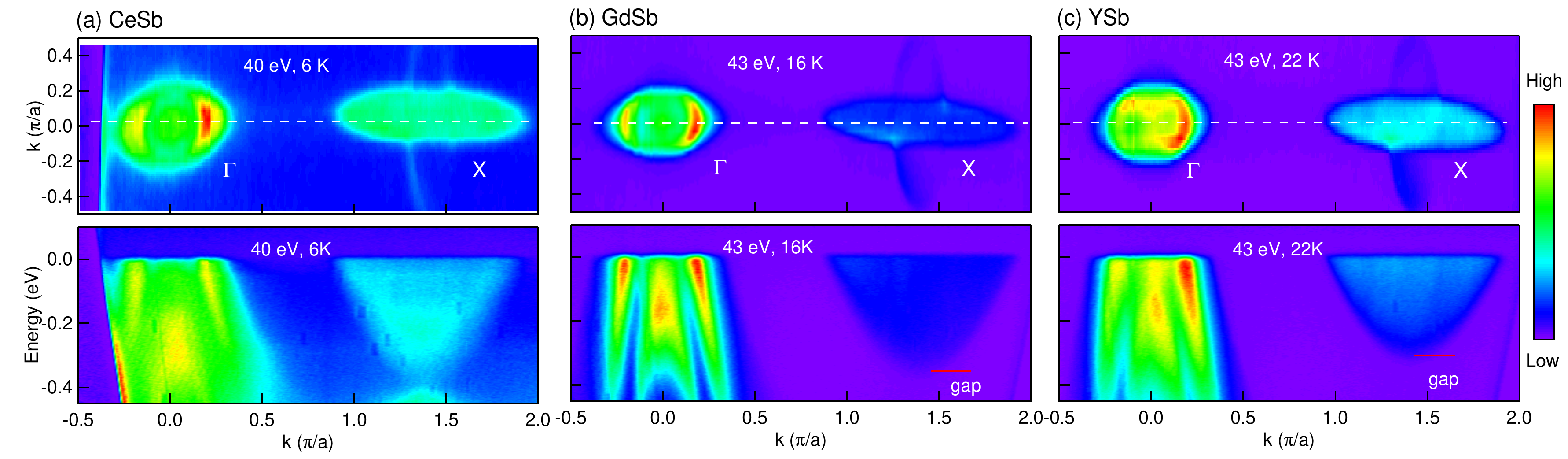}
	\caption{FS and high symmetry cuts (along the white dash lines in each FS plot) of YSb, CeSb, and GdSb. 
	(a) FS and band dispersion of YSb measured using the photon energy of 43~eV and $T=$22~K.
	(b) FS and band dispersion of CeSb measured using the photon energy of 40~eV and $T=$6~K.
	(c) FS and band dispersion of GdSb measured using the photon energy of 43~eV and $T=$16~K.
	\label{fig:Fig7}}
\end{figure*}

\begin{figure*}[bt]
	\includegraphics[width=6in]{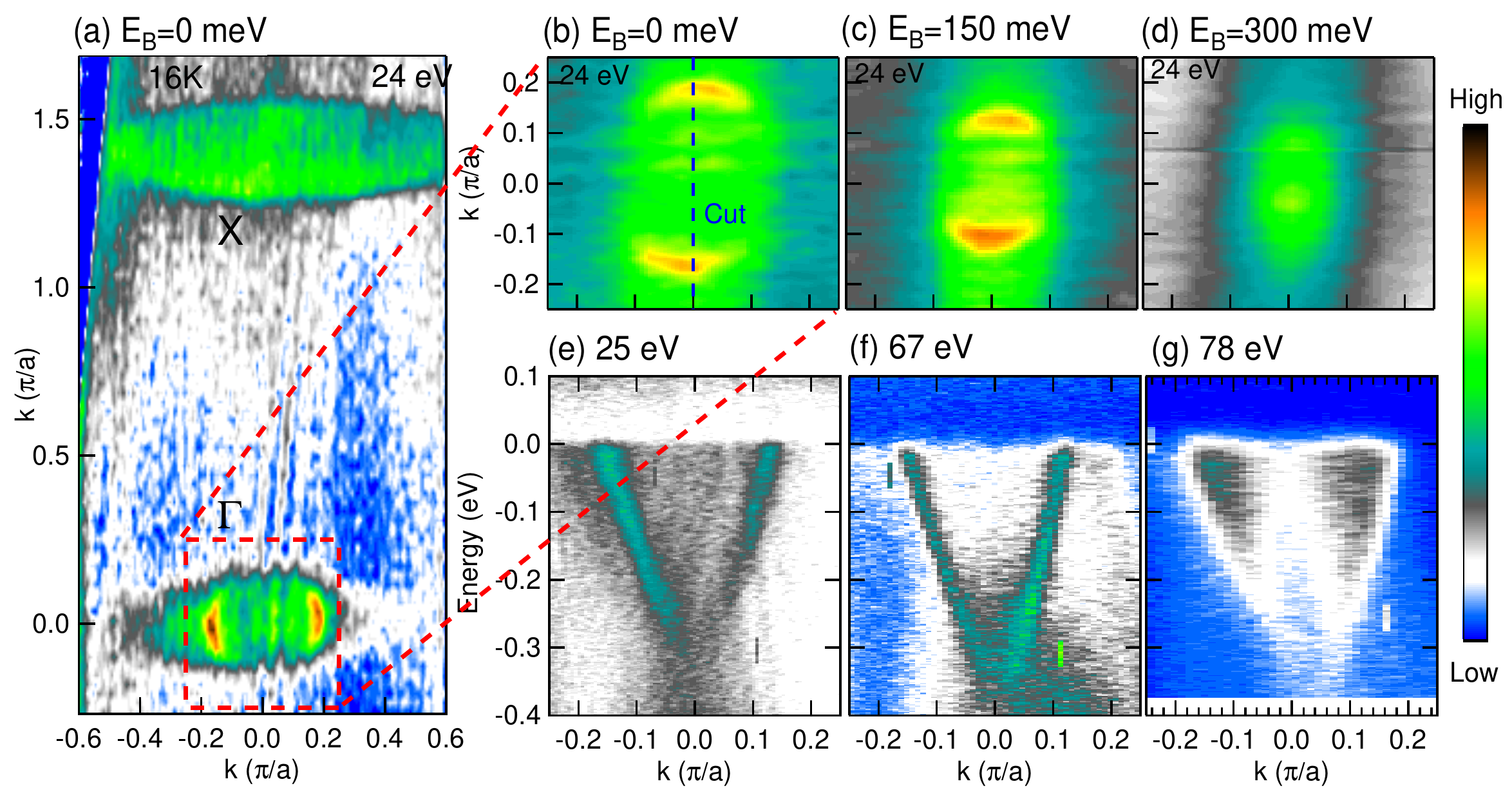}
	\caption{FS and high symmetry cuts of GdSb. 
	(a) FS of GdSb measured using the photon energy of 24~eV. 
	(b) FS (rotated ${90}^{\circ}$) of the zoom in area in (a).
	(c) constant energy contour plot at the binding energy of 150~meV.
	(d) constant energy contour plot at the binding energy of 300~meV.
	(e) Band dispersion along the cut as shown in (b) measured using the photon energy of 25~eV.
	(f) The same cut as in (e) measured using the photon energy of 67~eV.
	(g) The same cut as in (e) measured using the photon energy of 78~eV.
	\label{fig:Fig8}}
\end{figure*}

Single crystals of \textit{R}Sb were grown either from a tin-rich ternary melt~\cite{Canfield92PMPB} for the light rare earths, or from an antimony-rich binary solution for the heavy rare earths~\cite{Canfield2001High}.  In all cases, high purity elements were placed into an alumina crucible which itself was sealed into an amorphous silica ampoule, heated to above 1000 ${}^{\circ}$C and then slowly cooled to a decanting temperature at which point the ampoule was placed into a centrifuge and excess solution was removed from the crystals~\cite{Canfield92PMPB, Canfield2001High}. We present a summary of key physical properties of \textit{R}Sb crystals in Table~\ref{tab:Tab1}. These data nicely demonstrate the lanthanide contraction effect.

The Full-potential Linear Augmented Plane Wave (FPLAPW) method ~\cite{Blaha2001WIEN2k} with the generalized gradient approximation (GGA)~\cite{Perdew1996Generalized} was used to calculate the theoretical FS. The spin-orbit interaction was included. To obtain self-consistent charge density, we employed ${R}_{MT} \times {k}_{max} = 8.0$ with muffin tin (MT) radii of 2.8, 2.7 a.u. for Lu, and Sb respectively. 315 $k$-points were selected in the irreducible Brillouin zone and calculations were iterated to reach the total energy convergence criterion which was 0.01~mRy/cell. For Fermi surface calculations, we divided the $-2\pi/a < {k}_{x}, {k}_{y} < 2\pi/a$ range of the ${k}_{x}$, ${k}_{y}$ planes with different ${k}_{z}$ values by 200$\times$200 mesh. Fig.2(a) (below) is the result of ${k}_{z} = 0.0~2\pi/c$. Since it is convenient to compare to experiment results we have used a reduced unit cell (a = b = 4.285~\AA, c = 6.060~\AA) for calculations.
 
Fermi Surfaces of YSb, CeSb, GdSb, and LuSb were measured at the Advanced Light Source (ALS) synchrotron based ARPES system, utilizing a Scienta SES2002 electron analyser. Momentum and energy resolutions were set at 0.014~\AA$^{-1}$ along the direction of the analyzer slits and 17~meV, respectively. The samples were cleaved at temperatures around 20~K, and kept at their cleaving temperatures throughout the measurements. ${k}_{z}$ measurements of YSb and CeSb were carried out at the Synchrotron Radiation Center (SRC) at Wisconsin with ARPES system consisting of R4000 electron analyzer. Detailed ${k}_{z}$ mappings of YSb, DySb, HoSb, TmSb, and LuSb were performed using a tunable VUV laser ARPES system, consisting of a Scienta R8000 electron analyzer, a picosecond Ti:Sapphire oscillator, and a fourth harmonic generator~\cite{Jiang14RSI}. Samples were cleaved \textit{in situ} at 40~K under ultrahigh vacuum (UHV), and kept at their cleaving temperatures throughout the measurements. Data were collected with tunable photon energies in the 5.3 to 6.7~eV range. Momentum and energy resolutions were set at $\sim$ 0.005~\AA$^{-1}$ and 1~meV, respectively. The size of the photon beam on the sample was $\sim$ 30~$\mu$m.

\section{Experimental results}

In Figs.~\ref{fig:Fig1}(a)--(d), we show the Fermi surface intensity plots of \textit{R}Sb (\textit{R} = Ce, Gd, Y, Lu) integrated within 10~meV about the chemical potential measured at the corresponding temperatures and photon energies as marked at the top left and right corners of each plot. In panel (a), we can see that there are at least two pockets at the $\Gamma$ point in CeSb, however, we can not resolve these two pockets with confidence in other rare-earth compounds as shown in panels~(b), (c), and (d). At the $X$ point, two elongated electron pockets can be clearly seen in Figs.~\ref{fig:Fig1}(a), (b), and (c). In panel~(d), the FS of LuSb clearly shows four fold symmetry, consistent with the simple cubic structure of the compound, although the relative intensity of each electron pockets varies due to the matrix elements. The structure of the Fermi surface for these different compounds are quite similar, with at least two hole pockets at the center and two electron pockets at each corner of the Brillouin Zone. These results confirm that the increased number of 4\textit{f} electrons in the rare-earth elements does not have a significant effect on the electronic structure of the \textit{R}Sb system near ${E}_{F}$. Thus the 4\textit{f} electrons are likely strongly localized, and shielded by the completely filled 5\textit{s}${}^{2}$, 5\textit{p}${}^{6}$ and 6\textit{s}${}^{2}$ shells. However, the differences in the other aspects of the band structure are also obvious. The size of the pockets in these compounds seems to be different, which may be due to the differences in the chemical potential. However, no solid conclusion can be drawn from this set of data since they are FS sheets measured at different ${k}_{z}$ as marked by the red dashed lines in Fig.~\ref{fig:Fig4}. More detailed results and analysis will be provided in the laser ARPES measurements as discussed below using much higher energy and momentum resolutions. In panels~(d1)--(d4), we show the constant energy contour plots of LuSb at the binding energies of 0.3, 0.5, 0.9, and 1.5~eV. As we move down from chemical potential [panel (d)] to the binding energy of 0.3~eV [panel~(d1)], we can clearly see that the constant energy intensity contours at the $\Gamma$ point expand, and the ones at the $X$ point shrink, demonstrating the hole and electron character of the Fermi pockets at the $\Gamma$ and $X$ point, respectively. In panel~(d2), an additional band with circular constant energy contour is detected at the $\Gamma$ point and the electron pockets at the $X$ point completely vanish. As we move further down to 0.9~eV, the constant energy contour at the $\Gamma$ point continue to expand and a new feature is detected at the $X$ point. Panel~(d4) shows that the constant energy contour at high binding energy is rather complex yet still highly symmetric with four-fold symmetry. 

Figs.~\ref{fig:Fig2}(a)--(c) and (d)--(f) show the calculated Fermi surface and band structure of YSb, GdSb, and LuSb. Similar Fermi surface and band structure clearly can be seen across these crystals. However, the chemical potential is different as shown in Fig.~\ref{fig:Fig2}(g). Minor differences in the chemical potentials can be seen between YSb and GdSb, which have similar lattice constants. On the other hand, GdSb and LuSb show significant differences in chemical potential. Later, we also will show ultrahigh resolution laser ARPES measurements to demonstrated this in Fig.~\ref{fig:Fig5}. 

Fig.~\ref{fig:Fig3}(a) shows the calculated FS of \textit{R}Sb at the chemical potential (${k}_{z} = 0.0~2\pi/c$), with two circular pockets and two squarish pockets at the $\Gamma$ point. At the $X$ point, two elongated pockets intersecting each other, similar to the nodal ring structures that were proposed to exist in lanthanum monopnictides~\cite{Zeng2015Topological}. Panel (b) shows the FS of LuSb measured using the photon energy of 88~eV, which matches relatively well calculated FS shown in Fig.~\ref{fig:Fig3}(a). The other elongated electron pocket at the $X$ point is not clearly visible in LuSb, most likely due to the effect of matrix elements. On the other hand, those intersecting elongated electron pockets in CeSb, GdSb, and YSb can be clearly seen in Figs.~\ref{fig:Fig1}(a)--(c). Panels~(c)--(f) show the ARPES intensity along the red dashed lines in Fig.~\ref{fig:Fig3}(b). The corresponding Momentum Dispersion Curves (MDCs) at the chemical potential are shown at the top subpanels of the ARPES intensity plot, with green arrows pointing to the peak positions of each visible Fermi crossing. Cut~\#1 illustrates the cut along $\Gamma-X-\Gamma$ direction. At least one electron pocket is clearly seen at the center ($X$ point) and two hole pockets at the edge ($\Gamma$ point) of the plot. Four peaks (corresponding to Fermi crossings) at the $X$ point can be seen in the MDCs in the top subpanel~(c), demonstrating that there are two electron pockets at the $X$ point. The red dashed lines are the results of the band structure calculations using FPLAPW method, which matches relatively well with the ARPES measurements. We should note that there are three hole bands and one electron band crossing the Fermi level at the $\Gamma$ point in the band structure calculations. However, in most of the compounds that we have measured, only two hole pockets are most often visible, possibly due to the off center ${k}_{z}$ positions. As shown in Fig.~\ref{fig:Fig3}(a), the calculated FS has ${k}_{z} = 0.0~2\pi/c$. Whereas, the Fermi surfaces shown in Figs.~\ref{fig:Fig1}(a)--(d) have ${k}_{z}$ values marked by the red dashed lines in Figs.~\ref{fig:Fig4}(a)--(d).

To determine the three-dimensionality of the electronic structure of \textit{R}Sb~\cite{Hufner03Springer}, it is essential to tune the incident photon energies which tunes the out of plane (${k}_{z}$) momentum. At ALS and SRC synchrotron light sources, we measured the band dispersion along ${k}_{z}$ direction using photon energies in the 20 to 150~eV range (Fig.~\ref{fig:Fig4}). Although the Fermi surfaces of these compounds show some similarity as shown in Fig.~\ref{fig:Fig1}, the ${k}_{z}$ dispersions shown in Fig.~\ref{fig:Fig4} display significant variations in intensity and shape. The size (cross section area) of the Fermi surface sheets from different materials shown in Fig.~\ref{fig:Fig4} also varies. LuSb shows the largest Fermi surface sheets (spans from -0.5 to 0.5~$\pi/a$) and CeSb the smallest (spans from -0.25 to 0.25~$\pi/a$). In panels~3(d1)--(d4), the band structure of LuSb measured using different photon energies show the three dimensional character of this compound. The corresponding ${k}_{z}$ values of 3(d1)--(d4) are marked using red dashed lines in Fig.~\ref{fig:Fig4}(d). At 85 and 105~eV [panels~(d1) and (d2)], only one hole pocket can be easily identified. However, at 125 and 145~eV [panels~(d3) and (d4)], four band crossings (i.e., two hole pockets) can be rather easily observed. Similar structures can be seen in all ${k}_{z}$ dispersion plots.

In order to get more detailed information about the ${k}_{z}$ dispersion in these compounds, we have utilized the ultrahigh resolution, tunable Laser ARPES system. We should note that due to the limited range of accessible photon energies in our laboratory-based laser source, we can only map out a portion of the Brillouin zone along the ${k}_{z}$ direction. Fig.~\ref{fig:Fig5}(a) shows the ARPES intensity of YSb close to the $\Gamma$ point measured using various photon energies from 5.54 to 6.36~eV. The band dispersion clearly shifts upwards as incident photon energies are lowered, and touches the Fermi level at the incident photon energy of 5.9~eV. At the photon energy of 5.54~eV, clear Fermi crossings are observed. Thus, the hole band close to the $\Gamma$ point in YSb reveals expected strong three-dimensionality. The band dispersions in DySb, HoSb, TmSb, and LuSb all have similar structures, as shown in panel~(b), except it seems that the chemical potential varies slightly for different rare-earth elements. For example, it appears that the value of the chemical potential is higher in LuSb, than DySb. The shift of ${E}_{F}$ is probably due to the difference in the size of the rare-earth ions, i.e., lanthanide contraction~\cite{Taylor72CH}, since all the partially filled 4\textit{f} electrons can be considered as part of the core and do not contribute much to the conduction bands of these materials. Therefore, smaller lattice constants will result in higher chemical potential. This is consistent with the electronic structure calculation results shown in Fig.~\ref{fig:Fig2}(g). Panel~(c) presents the MDCs of the corresponding materials measured using the specific photon energies at ${E}_{F}$. The red dots mark the peak positions of the MDCs obtained by using double Lorentzian function fits. These data clearly show that the size of the FS strongly depends on the photon energy, thus ${k}_{z}$. By collecting the data for all these compounds using various photon energies, we successfully determined their ${k}_{z}$ dispersion shown in panel~(d). The red solid dots in panel~(d) represent the peak positions of the MDCs as shown in panel~(c). The blue dashed lines are guides to the eye and clearly reveal shapes of the Fermi surface along the ${k}_{z}$ direction. 

As previously discussed, CeSb and GdSb have large magnetoresistance. Thus we will discuss the electronic structure of these compounds in more details (YSb was found to have large magnetoresistance recently~\cite{Ghimire2016Magnetotransport, Yu2016Magnetoresistance, Pavlosiuk2016Giant}). As it was demonstrated in Fig.~\ref{fig:Fig3}(c), the two intersecting electron pockets at the $X$ point can not be easily resolved from the band dispersion in LuSb. To demonstrate that there are indeed two electron pockets at the $X$ point, we have plotted the enhanced diagram around the $X$ point for GdSb that was measured using the photon energy of 30~eV in Fig.~\ref{fig:Fig6}. Panels~(b) and (c) show the band dispersion along cuts~\#1 and 2. In Fig.~\ref{fig:Fig6}(b), two electron pockets can be clearly seen, showing a ``W''-like shape. However, the last part of ``W'' is not very visible probably due to the matrix elements effect. Panel~(c) presents the band dispersion along Cut~\#2 in (a), and demonstrates that the two electron bands in panel (b) are degenerate, with a visible gap between the upper and lower bands. These results confirm that there are indeed two electron pockets at the $X$ point with a gap between the conduction and valance bands. Upon the completion of this work, we noticed that similar structure is also reported in CeSb, where  new type of four-fold degenerate fermions were proposed~\cite{Alidoust2016New}. 

To answer the question that whether other rare-earth monoantimonides host such four-fold degenerate states, we have plotted the FS and high symmetry cuts along $\Gamma-X$ direction from YSb, CeSb, and GdSb, in Fig.~\ref{fig:Fig7}. In CeSb [panel (b)], the two intersecting electron bands seem to be touching with the top of the lower bands, consistent with the results from Ref.~\onlinecite{Alidoust2016New}. However, in YSb and GdSb, there is a significant gap (so significant that we cannot see the lower band in this energy range) between the electron pockets and the lower hole bands as seen from the band dispersion along the high symmetry cuts. 

What is interesting is that similar to the results in LaBi~\cite{Wu2016Asymmetric}, we have also observed a Dirac-like electron band at the $\Gamma$ point in YSb, CeSb, and GdSb for some specific values of photon energy. The details of the Dirac-like band in GdSb are shown in Fig.~\ref{fig:Fig8}. Panel~(a) shows the FS of GdSb measured using the photon energy of 24~eV. The zoom in image of the red box in (a) is presented in (b). Panel~(c) shows the constant energy contour plot at the binding energy of 150~meV, where we can still recognize the circular shape. The constant energy contour plot at the binding energy of 300~meV is shown in (d), showing the electron pocket shrinks down to a single point at the center. Panels~(e)--(g) show the high symmetry cut [as marked in (b)] measured using some photon energies, which shows rather linear dispersive bands. This Dirac-like structure may contribute to the unusually high magnetoresistance in these materials~\cite{Kasuya96JPSJ, Li96PRB, Ghimire2016Magnetotransport, Yu2016Magnetoresistance, Pavlosiuk2016Giant}.

\section{Conclusion}
The \textit{R}Sb family is an ideal system for studying the evolution of electronic structure due to different rare-earth ions. We successfully measured the FS of rare-earth monoantimonides with \textit{R} = Y, Ce, Gd, Dy, Tm, Ho, and Lu, by using synchrotron radiation and laser-based ARPES systems. Fermi surfaces of different materials measured using different photon energies show similar structure of at least two hole pockets centered at the $\Gamma$ point and two intersecting electron pockets at the $X$ points. The results match relatively well with the band structure calculations. By using the synchrotron and tunable VUV laser ARPES systems, we mapped the ${k}_{z}$ dispersion of \textit{R}Sb and concluded that the inner hole band centered at the $\Gamma$ point have strong three-dimensionality. By comparing the band structure for different rare earth elements, we show that the 4\textit{f} electrons in these ions do not qualitatively affect the electronic structure close to the chemical potential. The ion size (because of Lanthanide contraction), on the other hand, has a significant effect on the chemical potential in these materials. With smaller crystal lattice, the chemical potential moves higher. The lanthanide contraction effect has been demonstrated by both ARPES measurements and electronic structure calculations. Though our instrumentation has limited our ability to probe the low temperature phase transitions of DySb and CeSb, our results provide an insight into the basic electronic structures of these materials. Further research is needed, especially measurements carried out at lower temperatures to study the magnetic phase transitions in these materials. 

\begin{acknowledgments}
We would like to thank Xin Zhao for helpful discussions. This work was supported by the U.S. Department of Energy (DOE), Office of Science, Basic Energy Sciences, Materials Science and Engineering Division. The research was performed at the Ames Laboratory, which is operated for the U.S. DOE by Iowa State University under contract \# DE-AC02-07CH11358. Y.W. (Analysis of ARPES data) was supported by Ames Laboratory’s Laboratory-Directed Research and Development (LDRD) funding. L.H. was supported by CEM, a NSF MRSEC, under Grant No. DMR-1420451.
\end{acknowledgments}

Raw data for this manuscript is available at \url{http://lib.dr.iastate.edu/ameslab_datasets/}.

\bibliography{RSb}

\end{document}